\documentstyle[twoside,fleqn,espcrc2,epsf]{article}

\title{
Coupling Constants for Scalar Glueball Decay\thanks{Talk
presented by D.~Weingarten}}

\author{
	J.~Sexton\thanks{Permanent address: Department of
Mathematics, Trinity College, Dublin 2, Republic of Ireland},
A.~Vaccarino\thanks{Present address: Piazza Giovanetti 1,
Novara, 28100 Italy}
and D.~Weingarten\\ 	IBM Research, P.O.~Box 218,
Yorktown Heights, NY 10598\\ }

\begin{document}

\begin{abstract}

We evaluate the partial decay widths of the lightest scalar glueball to
pairs of pseudoscalar quark-antiquark states.  The calculation is done
in the valence (quenched) approximation on a $16^3 \time 24$ lattice at
$\beta = 5.7$. These predictions and values obtained earlier for the
infinite volume continuum limit of the scalar glueball's mass are in
good agreement with the observed properties of $f_J(1710)$ and
inconsistent with all other observed meson resonances.

\end{abstract}

\maketitle

\section{Introduction}

Whether or not glueballs have been observed in experiment is still
generally considered unsettled.  Since the properties of glueballs are
not expected to be drastically different from the properties of flavor
singlet bosons including valence quarks and antiquarks, the
identification in experiment of states with large glueball contributions
is difficult without a reliable calculation of the predictions of QCD.
The lattice formulation of QCD appears to us to give the most reliable
method now available for determining QCD's predictions for the masses
and decay couplings of glueballs.

For the infinite volume continuum limit of the valence (quenched)
approximation to the lightest scalar glueball mass, we reported some
time ago \cite{Vaccarino} the value 1740(71) MeV.  This prediction was
obtained using ensembles of 25000 to 30000 gauge configurations on each
of several different lattices.  An earlier independent valence
approximation calculation \cite{Livertal} of the scalar glueball mass,
extrapolated to the continuum limit \cite{Weingarten94} following
Ref.~\cite{Vaccarino}, yields a prediction of 1625(94) MeV.  The
calculation of Ref~\cite{Livertal} uses ensembles of between 1000 and
3000 configurations on several different lattices.  Combining the two
mass calculations and taking into account the correlation between their
statistical uncertainties arising from a common procedure for converting
lattice quantities into physical units gives a scalar glueball mass of
1707(64) MeV. This result and the mass prediction with larger
statistical weight are both in good agreement with the mass of
$f_J(1710)$ and are strongly inconsistent with all but
$f_0(1500)$~\cite{Amsler1} among the other established flavor singlet
resonances which could be scalars.  For $f_0(1500)$ the disagreement is
still by more than three standard deviations.

The valence approximation, in effect, replaces the momentum and
frequency dependent color dielectric constant arising from
quark-antiquark vacuum polarization with its zero-momentum,
zero-frequency limit \cite{Weingarten82}. For the long distance
properties of hadrons, we would expect this approximation to be fairly
reliable.  The infinite volume continuum limits, for example, of the
valence approximation to the masses of eight low lying hadrons composed
of quarks and antiquarks differ from experiment by amounts ranging up to
6\% \cite{Butler}.  A 6\% error in the glueball mass would be 100 MeV.
An adaptation of an argument giving a negative sign for the valence
approximation error in $f_{\pi}$~\cite{Butler} also suggests a negative
sign for the glueball mass error.  Thus we would expect the scalar
glueball in full QCD to lie between 1707(64) MeV and 1807(64) MeV, again
favoring $f_J(1710)$ with $f_0(1500)$ still possible but improbable.

The key question not answered by the mass results, however, is whether
the lightest scalar glueball has a decay width small enough for this
particle actually to be identified in experiment. It seems likely to us
that a scalar glueball with a width of a few hundred MeV or less and
mass in the neighborhood of 1700 MeV would already have been seen in
experiment. Alternatively, if the scalar glueball has a width of a GeV
or more, the prospect of ever finding this particle seems remote.  A
further question in the identification of $f_J(1710)$ as a glueball is
raised by the argument that since glueballs are flavor singlets they
should have the same couplings to $2 \pi_0$, to $2 K_L$, and to $2
\eta$. This equality is violated by $f_J(1710)$ decay couplings.

\begin{figure}
\begin{center}
\vskip -5mm
\leavevmode
\epsfxsize=65mm
\epsfbox{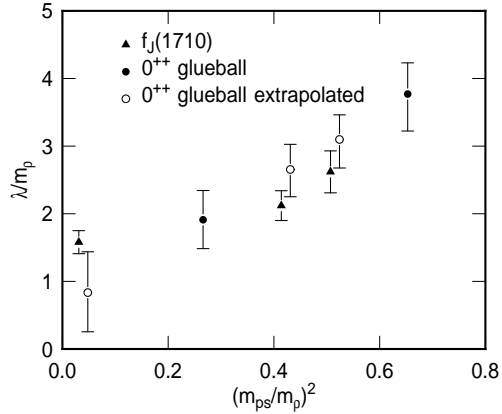}
\vskip -13mm
\end{center}
\caption{ 
Couplings for decay to a pair of pseudoscalars as a function
of pseudoscalar mass squared.}
\label{fig:lambdas}
\vskip -9mm
\end{figure}

In the present article we report the first lattice QCD calculation of
the valence (quenched) approximation to the decay couplings of the
lightest scalar glueball to pairs of pseudoscalar quark-antiquark
states.  The couplings constants we obtain, combined with the mass
prediction of 1740(71) MeV, give a total two-pseudoscalar decay width of
108(29) MeV.  With any reasonable guess concerning the scalar glueball's
branching fraction to multibody decay modes, the resulting total decay
width is well below 200 MeV and therefore small enough for the scalar
glueball to be identified in experiment. In fact, the predicted total
two-pseudoscalar decay width, and individual couplings to $2 \pi_0$, to
$2 K_L$, and to $2 \eta$ are all in good agreement with properties of
$f_J(1710)$ and inconsistent with all other established flavor singlet
resonances which could be scalars.  The total two-body width of
$f_J(1710)$ is 99(15) MeV
\cite{Lind}.  A comparison of our predicted coupling constants with data
for $f_J(1710)$ \cite{Lind} is shown in Figure~\ref{fig:lambdas}.

The calculation we present uses 10500 independent gauge configurations
on a lattice $16^3 \times 24$ at $\beta = 5.70$. The corresponding
inverse lattice spacing is 1.35 GeV. We believe this lattice has spacing
sufficiently small and volume sufficiently large to give partial widths
within 30\% of their infinite volume continuum limits.

In the valence approximation, according to one point of view, glueballs
are pure field and are not mixed with states including valence quarks or
antiquarks.  The agreement between the glueball mass and decay couplings
found in the valence approximation and the observed mass and decay
couplings of $f_J(1710)$ appears to us to be strong evidence that this
state is largely a scalar glueball with at most some relatively smaller
amplitude for configurations including valence quark-antiquark pairs.

The calculations presented here were carried out on the GF11 parallel
computer \cite{Weingarten90} at IBM Research and took approximately two
years to complete at a sustained computation rate of between 6 and 7
Gflops. A preliminary version of this work is discussed in
Ref.~\cite{Sexton95}.

In the remainder of this paper we describe our method for determining
scalar glueball decay couplings then present our numerical results.

\section{Method}

We work with a euclidean lattice gauge theory on a lattice $L^3 \times
T$, with the plaquette action for the gauge field and the Wilson action
for quarks. We assume initially exact flavor SU(3) symmetry for the
quark mass matrix.  Each gauge configuration is fixed to Coulomb gauge.
We then define a collection of smeared fields.  We describe smearing
only for the particular choice of parameters actually used in the decay
calculation.  Let $U_i(x)$ for a space direction $i = 1, 2, 3,$ be a
smeared link field~\cite{Vaccarino} given by the average of the 9 links
in direction $i$ from the sites of the (3 site) x (3 site) square
oriented in the two positive space directions orthogonal to $i$ starting
at site $x$. Let $V_{ij}(x)$ be the trace of the product of $U_i(x)$ and
$U_j(x)$ around the outside of a (3 link) x (3 link) square.  Let
$g(t)$, carrying zero momentum, be the sum of the $V_{ij}(x)$ for all
$i, j$ and $x$ with time component $t$.  Define $\overline{\Psi}(x)$ and
$\Psi(x)$ to be quark and antiquark fields, respectively,
smeared~\cite{Butler} by convoluting the local fields with a space
direction gaussian with mean-square radius 6.0. The smeared field
$\pi_i(x)$ with flavor index $i$ is $\overline{\Psi}(x) \gamma^5
\Lambda_i \Psi(x)$, where $\Lambda_i$ is a Gell-Mann flavor matrix.
Define $\tilde{\pi}_i(\vec{k},t)$ to be the Fourier transform of
$\pi_i(x)$ on the time $t$ lattice hyperplane.

$E^{\pi}_1$ and $E^{\pi}_2$ are defined as the energy of a single
pseudoscalar at rest or with momentum magnitude $|\vec{k}| = 2 \pi/ L$,
respectively. The field strength renormalization constant $\eta^{\pi}_1$
is defined by setting the large $t$ asymptote of the vacuum expectation
value $< \tilde{\pi}^{\dagger}_i( 0,t) \tilde{\pi}_i( 0,0)>$ to
$(\eta^{\pi}_1)^2 L^3 exp( -E^{\pi}_1 t)$. Define $\eta^{\pi}_2$ similarly
from a pseudoscalar field with momentum magnitude $|\vec{k}| = 2 \pi/
L$.  The glueball mass $E^g$ and field strength renormalization constant
$\eta^g$ are defined by setting the large $t$ asymptote of $<g(t)
g(0))>$ to be $(\eta^g)^2 L^3 exp( -E^g t)$.

Let the flavor singlet, rotationally invariant, two-pseudoscalar field
$\Pi( t_1, t_2)$ be $(16)^{-1/2} \sum_i \pi_i(0,t_1) \pi_i(0,t_2)$, with
the sum over $i$ running from 1 to 8.  Define the flavor singlet field
for two zero momentum pseudoscalars $\tilde{\Pi}_1( t_1, t_2)$ as
$(16)^{-1/2} \sum_i \tilde{\pi}_i(0,t_1)
\tilde{\pi}_i(0,t_2)$. Let the field $\tilde{\Pi}_2( t_1, t_2)$
for two pseudoscalars with opposite momenta be $(24)^{-1/2}
\sum_{i \vec{k}} \tilde{\pi}_i(\vec{k},t_1) \tilde{\pi}_i(-\vec{k},t_2)$ 
where the sum for $\vec{k}$ is over the three positive orientations with
$|\vec{k}| = 2 \pi/ L$.

Define $| 1>$ and $| 2>$, respectively, to be the lowest and second
lowest energy flavor singlet, rotationally invariant two-pseudoscalar
states, both normalized to 1.  $E^{\pi\pi}_i$ is the energy of $| i>$.
The amplitude $\eta^{\pi \pi}_{ij}(t)$ is then defined as $L^{-3} < i|
\tilde{\Pi}_j( t, 0) | \Omega >$. At large $t$, $\eta^{\pi
\pi}_{ij}(t)$ approachs $\eta^{\pi \pi}_{ij}
exp(-E^{\pi}_j t)$.  We expect the diagonal coefficients $\eta^{\pi
\pi}_{11}$ and $\eta^{\pi \pi}_{22}$ to be larger than the off-diagonal
$\eta^{\pi \pi}_{21}$ and $\eta^{\pi \pi}_{12}$, respectively. Since
pairs of pseudoscalars can interact and scatter, however, the
off-diagonal coefficients need not be zero.

The connected three-point functions $T_i( t_g, t_{\pi})$ from which we
extract coupling constants are $< g(t_g) \tilde{\Pi}_i( t_{\pi}, 0) > -
< g(t_g)> < \tilde{\Pi}_i( t_{\pi}, 0) >$.  If the hopping constant
$\kappa$ is chosen so that $E^{\pi\pi}_1 = E^g$, the lightest
intermediate state which can appear between the glueball and
pseudoscalars in $T_1( t_g, t_{\pi})$ is $| 1>$. For large $t_g$ with
$t_{\pi}$ fixed, $T_1( t_g, t_{\pi})$ will therefore be proportional to
the coupling constant of a glueball to two pseudoscalars at rest.  If
the $\kappa$ is chosen so that $E^{\pi\pi}_2 = E^g$, however, the
lightest intermediate state which can appear between the glueball and
pseudoscalars in $T_2( t_g, t_{\pi})$ is still $| 1>$, not $| 2>$,
barring the unlikely occurrence of $\eta^{\pi \pi}_{1 2}(t) = 0$.  To
get the coupling of a glueball to two pseudoscalars with momenta of
magnitude $2
\pi L^{-1}$ from $T_2( t_g, t_{\pi})$, the contribution to $T_2( t_g,
t_{\pi})$ from the $| 1>$ intermediate state must be removed.

We therefore define the subtracted amplitudes
\begin{eqnarray}
\label{defs}
S_i( t_g, t_{\pi})  = T_i( t_g, t_{\pi})  -
\frac{\eta^{\pi \pi}_{ji}(t_{\pi})} 
{\eta^{\pi \pi}_{jj}(t_{\pi})} T_j( t_g, t_{\pi}), 
\end{eqnarray}
for $(i,j)$ of either (1,2) or (2,1).  In $S_i( t_g, t_{\pi})$ the
contribution from the intermediate state $| j>$ has been canceled.
Although the subtraction in $S_1( t_g, t_{\pi})$ is irrelevant for large
enough $t_g$, we expect the subtracted $S_1( t_g,t_{\pi})$ to approach
its large $t_g$ behavior more rapidly than $T_1( t_g, t_{\pi})$
approaches its large $t_g$ limit.

Another state which can also appear between the pseudoscalars and
glueball in $T_i( t_g, t_{\pi})$ consists of a quark and an antiquark
bound as a scalar flavor singlet.  For the lattice size, $\beta$ and
$\kappa$ used in the present calculation, this state we have found 
has a mass in lattice units above 1.25. The scalar glueball mass is
0.972(44).  The scalar quark-antiquark state therefore will make only
its appropriate virtual contribution and does not require an additional
correction. 

For $t_g$ and $t_{\pi}$ large with $(t_g + t_{\pi}) \ll T$, the three-point
functions become
\begin{eqnarray}
\label{Sasym}
S_i( t_g, t_{\pi}) \rightarrow 
\frac{c_i \sqrt{3} \lambda_i \eta^g \eta^{\pi \pi}_{ii}
(1 - r) L^3} {\sqrt{8 E^g (E^{\pi}_i)^2}} 
s_i(t_g, t_{\pi}).
\end{eqnarray}
Here $c_1 = 1/\sqrt{2}$, $c_2 = \sqrt{3}$, $r$ is $(\eta^{\pi \pi}_{12}
\eta^{\pi \pi}_{21}) / (\eta^{\pi \pi}_{11} \eta^{\pi \pi}_{22})$ and
$\lambda_1$ and $\lambda_2$ are the glueball coupling constants to a
pair of pseudoscalars at rest or with momenta of magnitude $2 \pi
L^{-1}$, respectively.  As discussed earlier, $\eta^{\pi \pi}_{ij}$ is defined
from $\eta^{\pi \pi}_{ij}(t)$ as large $t$.
The factor $s_i(t_g, t_{\pi})$ is
\begin{eqnarray}
\label{defsi}
s_i(t_g, t_{\pi}) & = &
\sum_t exp[ -E^g |t - t_g| - E^{\pi}_i |t| - \nonumber \\ 
& & E^{\pi}_i |t - t_{\pi}| - \delta_i( t, t_{\pi})|t - t_{\pi}|]. 
\end{eqnarray}
For $t \ge t_{\pi}$, $\delta_i(t, t_{\pi})$ is the binding energy
$E^{\pi \pi}_i - 2 E^{\pi}_i$ and otherwise $\delta_i(t, t_{\pi})$ is 0.

The coupling constants in Eq.~(\ref{Sasym}) have been identified by
fitting $S_i(t_g, t_{\pi})$ with the three-point function arising from
a simple phenomenological interaction lagrangian. This procedure for
determining $\lambda_i$ is correct to leading order in $\lambda_i$.
A corresponding relation has been used for some time to find coupling
constants among hadrons containing quarks and recently has produced a
variety of results in good agreement with experiment~\cite{decays}. The
normalization of the $\lambda_i$ is chosen so that in the
continuum limit they become, up to a factor of $-i$, Lorentz-invariant
decay amplitudes with the standard normalization convention used in the
section on kinematics of the Review of Particle Properties.

To extract $\lambda_i$ from $S_i( t_g, t_{\pi})$ using Eq.~(\ref{Sasym}) 
we need the $\eta^{\pi \pi}_{ij}$.  These we obtain from
propagators for two-pseudoscalar states.  Define 
$C_i(t_1,t_2)$ to be $< \Pi(t_1 + 2 t_2, t_1 + t_2)
\tilde{\Pi}_i( t_2, 0)>$.  
For moderately large values of $t_1$, we then have
\begin{eqnarray}
\label{Casym}
C_i(t_1,t_2) & = & C_{1i}(t_2) exp( -E^{\pi\pi}_1 t_1) + \\ & &
C_{2i}(t_2) exp( -E^{\pi\pi}_2 t_1) + D_i( t_2), \nonumber \\
\label{Cxasym}
C_{ij}(t_2)  & = &
\eta^{\pi \pi}_{i1}(t_2) \eta^{\pi \pi}_{ij}(t_2) + \\
& & \sqrt{6} \eta^{\pi \pi}_{i2}(t_2) \eta^{\pi \pi}_{ij}(t_2). \nonumber 
\end{eqnarray}
The term $D_i( t_2)$, independent of $t_1$, arises from propagation
across the lattice's periodic time direction boundary.  $D_i( t_2)$
makes a significant contribution to $C_i( t_1, t_2)$ only if $t_1 + t_2$
is comparable to $T$.  From Eqs.~(\ref{Casym}) and (\ref{Cxasym}),
$\eta^{\pi \pi}_{ij}(t)$ and $\eta^{\pi \pi}_{ij}$ can be found.

One of the jobs performed by the $\eta^{\pi \pi}_{ij}$ in
Eq.~(\ref{Sasym}) is to correct for the interaction between the two
pseudoscalars produced by a glueball decay. In the valence approximation
pseudoscalars interact without the production and annihilation of
virtual quark-antiquark pairs. The $C_i(t_1,t_2)$ from which the
$\eta^{\pi \pi}_{ij}$ are obtained should therefore be evaluated from
quark propagators including only terms in which all initial quarks and
antiquarks propagate through to final quarks and antiquarks.  It can be
shown that including also in the two-pseudoscalar propagator terms in
which initial quarks propagate to initial antiquarks and final quarks
propagate to final antiquark would lead to $\eta^{\pi \pi}_{ij}$ which
correct Eq.~(\ref{Sasym}) for processes missing from glueball decay in
the valence approximation.

If $t_1$ is made very large, the $C_i(t_1,t_2)$ are given by a sum of
two terms each of which is a slightly more complicated version of one of
the exponentials in Eq.~(\ref{Casym}).  This complication occurs because
in the valence approximation the exchange of a particle composed of
quarks and antiquarks between the pseudoscalars produced in a glueball
decay is not iterated in the same way as in full QCD.  The term in
Eq.~(\ref{Casym}) with coefficient $C_{ji}(t_2)$ requires no
modification if $| E^{\pi \pi}_j - 2 E^{\pi}_j |^2 t_1^2 / 2 \ll 1$.
Within the intervals of $t_1$ we use to determine the $\eta^{\pi
\pi}_{ij}$, these bounds are well satisfied. In any case,  
as we will discuss below, the values we obtain for $\eta^{\pi \pi}_{ij}$
turn out to be close to their values for noninteracting pseudoscalars.
A consequence is that the corrections due to interactions between the
decay pseudoscalars which the $\eta^{\pi \pi}_{ij}$ contribute to the
predicted values of $\lambda_i$ are small, and our results are fairly
insensitive to details of the two-pseudoscalar interaction.

\vskip -1mm

\begin{figure}
\begin{center}
\vskip -5mm
\leavevmode
\epsfxsize=65mm
\epsfbox{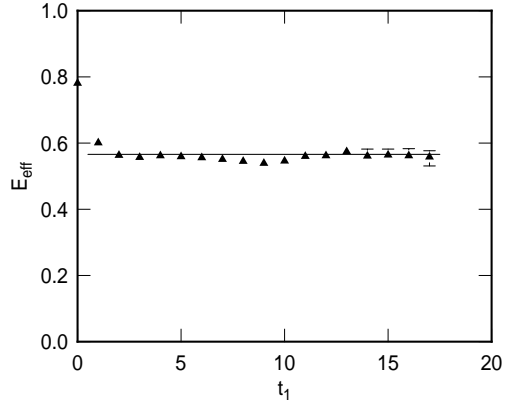}
\vskip -14mm
\end{center}
\caption{ 
Effective masses determined from the $t_1$ dependence
of $C_1( t_1, t_2) - D_1( t_2)$ for $\kappa = 0.1675$ and $t_2 = 2$.
}
\label{fig:mom0}
\vskip -9mm
\end{figure}

\begin{figure}
\begin{center}
\vskip -5mm
\leavevmode
\epsfxsize=65mm
\epsfbox{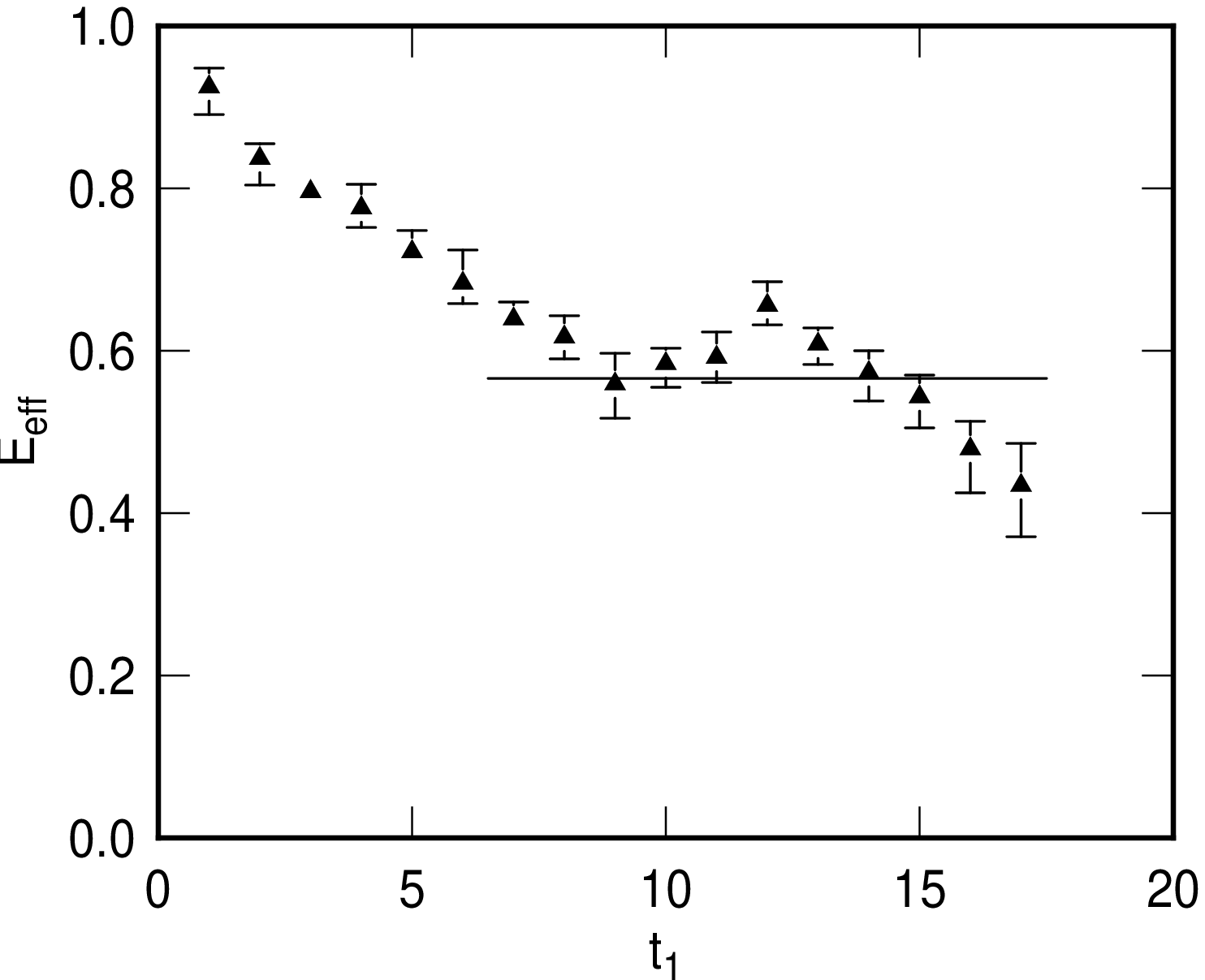}
\vskip -13mm
\end{center}
\caption{ 
Effective masses determined from the $t_1$ dependence
of $C_2( t_1, t_2) - D_2( t_2)$ for $\kappa = 0.1675$ and $t_2 = 2$.
}
\label{fig:mom1fit0}
\vskip -8mm
\end{figure}

\begin{figure}
\begin{center}
\vskip -5mm
\leavevmode
\epsfxsize=65mm
\epsfbox{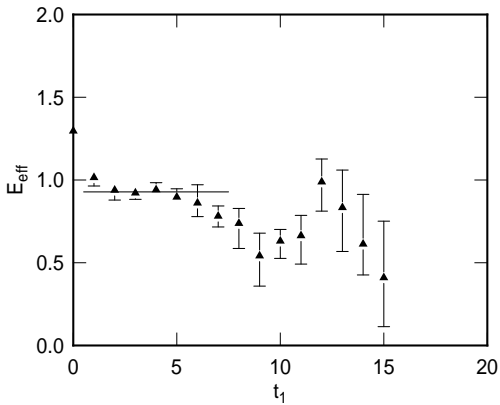}
\vskip -13mm
\end{center}
\caption{ 
Effective masses determined from the $t_1$ dependence
of $C_2( t_1, t_2) - C_{12}(t_2) exp( -E^{\pi\pi}_1 t_1) - D_2( t_2)$ 
for $\kappa = 0.1675$ and $t_2 = 2$.
}
\label{fig:mom1fit1}
\vskip -8mm
\end{figure}

\section{Results}

We now turn to our numerical results.  In all of our numerical work,
fits were done by minimizing $\chi^2$ found from the full correlation
matrix among the fitted data and statistical uncertainties were found by
the bootstrap method.

At $\beta = 5.7$ on a $16^3 \times 24$ lattice, with an ensemble of
10500 independent configurations, we determined glueball and single
pseudoscalar energies and renormalization constants following
Refs.~\cite{Vaccarino} and
\cite{Butler}, respectively. 
The energies in lattice units are shown in Table~\ref{tab:1particle}.

\begin{table}
\begin{center}
\begin{tabular}{ccl}
\hline
$\kappa$ &  & \\
\hline
0.1650 & $E^{\pi}_1$ & 0.45717(38) \\
       & $E^{\pi}_2$ & 0.59815(69) \\
\hline
0.1675 & $E^{\pi}_1$ & 0.28905(83) \\
       & $E^{\pi}_2$ & 0.48686(163) \\
\hline
$-$ & $E^g$ & 0.972(44) \\
\hline
\end{tabular}
\caption{Single particle energies in lattice units.}
\label{tab:1particle}
\end{center}
\vskip -15mm
\end{table}

On a lattice of size $16^3 \times 40$ we then evaluated the
two-pseudoscalar propagator $C_i( t_1, t_2)$ at $\kappa = 0.1650$ using
107 independent configuration, and at $\kappa = 0.1675$ using 875
independent configurations.  Fitting the $t_1$ dependence of
$C_i(t_1,t_2)$ to Eqs.~(\ref{Casym}) and (\ref{Cxasym}), we determined
$E^{\pi \pi}_i$ and $\eta^{\pi \pi}_{ij}(t_2)$ for a range of different
$t_2$.  For each value of $\kappa$ we first found $E^{\pi
\pi}_1$ by fitting the $t_1$ dependence of $C_1(t_1,t_2)$ with fixed
$t_2$ to Eq.~(\ref{Casym}). This fit was done at large enough $t_1$ for
the contribution of the more rapidly falling term $C_{21}(t_2) exp(
-E^{\pi\pi}_2 t_1)$ in Eq.~(\ref{Casym}) 
to be negligible. Since the coefficient $C_{21}(t_2)$ is
significantly smaller than $C_{11}(t_2)$, it is not difficult to find a range
of $t_1$ within which only the $C_{11}(t_2)$ term survives.
For both values of $\kappa$ we found
$11 - t_2 \le t_1 \le 18 - t_2$ to be satisfactory.
Figure~\ref{fig:mom0} shows the fit to $C_1(t_1,t_2)$ for $\kappa =
0.1675$, $t_2 = 2$. The horizontal line is the fitted value of $E^{\pi
\pi}_1$. The points with error bars are effective energies
$E_{eff}$ given by $\log[c_1(t_1, t_2)/c_1(t_1+1, t_2)]$ for $c_1(t_1,
t_2)$ defined as the propagator with constant term removed $C_1(t_1,t_2) -
D_1(t_2)$.

We then looked for a region at large $t_1$ in which $C_2(t_1,t_2)$ gets
negligible contribution from $C_{22}(t_2) exp( -E^{\pi\pi}_2 t_1)$ in
Eq.~(\ref{Casym}).  This region is more difficult to locate than the
corresponding region for $C_1(t_1,t_2)$, since the coefficient
$C_{12}(t_2)$ of the dominant large $t_1$ term in Eq.~(\ref{Casym}) is
smaller than the coefficient $C_{22}(t_2)$ of the term which falls more
rapidly at large $t_1$.  In these fits, we therefore took as input the
value already determined for $E^{\pi \pi}_1$ from the fit to
$C_1(t_1,t_2)$.  A satisfactory range for both $\kappa$ we found to be
$14 - t_2
\le t_1 \le 19 - t_2$.  For the fit to $C_2(t_1,t_2)$ with $\kappa = 0.1675$, $t_2 = 2$,
Figure~\ref{fig:mom1fit0} shows $E^{\pi \pi}_1$ in comparison to
effective energies determined from $c_2( t_1, t_2)$ defined as
$C_2(t_1,t_2) - D_2(t_2)$.  With $C_{12}(t_2)$ determined, we then formed the
subtracted propagator $d_2( t_1, t_2)$ given by $C_2(t_1,t_2) - C_{12}(t_2)
exp( -E^{\pi\pi}_1 t_1) - D_2(t_2)$ and looked for a plateau in
effective masses at smaller $t_1$ to be fit to $C_{22}(t_2) exp(
-E^{\pi\pi}_2 t_1)$.  A satisfactory range we found to be the 4
successive time values in the interval $2 \le t_1 \le 7$ giving the fit
with the smallest $\chi^2$ per degree of freedom.  For the fit to
$d_2(t_1,t_2)$ with $\kappa = 0.1675$, $t_2 = 2$,
Figure~\ref{fig:mom1fit1} shows $E^{\pi \pi}_2$ determined from $3 \le
t_1 \le 6$ in comparison to effective energies determined from $d_2(
t_1, t_2)$.

\begin{table}
\begin{center}
\begin{tabular}{cr@{}ll}
\hline
$\kappa$ & &  \\
\hline
0.1650 & \multicolumn{2}{c}{$E^{\pi \pi}_1$}  & 0.9076(49)  \\
       & \multicolumn{2}{c}{$E^{\pi \pi}_2$}  & 1.2180(155) \\
\hline
0.1675 & \multicolumn{2}{c}{$E^{\pi \pi}_1$}  & 0.5690(44)  \\
       & \multicolumn{2}{c}{$E^{\pi \pi}_2$}  & 0.8925(243) \\
\hline
       & \multicolumn{3}{c}{$\hat{\eta}^{\pi \pi}_{ij}$} \\
\hline
0.1650 & 0 & .988(30)    & 0.091(8) \\
       & $-0$ & .087(8) & 1.065(13) \\
\hline
0.1675 & 1 & .050(21)    & 0.107(6) \\
       & $-0$ & .112(8) & 1.053(53) \\
\hline
\end{tabular}
\caption{Energies in lattice units and field-strength 
renormalizations for two-pseudoscalar states.}
\label{tab:2particle}
\end{center}
\vskip -15mm
\end{table}

At $\kappa = 0.1650$ we obtained results for $0 \le t_2 \le 4$, and at
0.1675 we found results for $t_2$ of 0 and $2 \le t_2
\le 5$.  The values of $E^{\pi \pi}_i$ were statistically consistent with being
independent of $t_2$ in all cases.  The $\eta^{\pi \pi}_{ij}(t_2)$ were
consistent with the asymptotic form $\eta^{\pi \pi}_{ij} exp( -E^{\pi}_j
t_2)$ in all cases for $t_2 \ge 2$.  The final values of $E^{\pi \pi}_i$
and $\hat{\eta}^{\pi \pi}_{ij}$ used in finding the $\lambda_i$ were taken from
combined fits with $2 \le t_2 \le 4$ for $\kappa = 0.1650$ and with $2
\le t_2 \le 5$ for $\kappa = 0.1675$. 
Table~\ref{tab:2particle} gives the final $E^{\pi \pi}_i$ in lattice
units and
$\hat{\eta}^{\pi \pi}_{ij}$.  For noninteracting pseudoscalars
$\hat{\eta}^{\pi \pi}_{ij}$
is 1 for $i = j$ and 0 otherwise.  Our data is close to these values.
The final value of $\lambda_1$ is changed by less than 1 standard
deviation and the final $\lambda_2$ is changed by less than 2 standard
deviations if we ignore the determination of $\hat{\eta}^{\pi \pi}_{ij}$ and
simply use the the noninteracting values.

From our fitting procedure we were unable to obtain a satisfactory
signal for the coefficient $C_{21}(t_2)$ which enters the determination
of $\hat{\eta}^{\pi \pi}_{21}$ and therefore of $\lambda_1$.  Since the
other $\eta^{\pi \pi}_{ij}$ are close to their values for pairs of
noninteracting pseudoscalars, we took $\hat{\eta}^{\pi \pi}_{21}$ from
the relation $\hat{\eta}^{\pi \pi}_{21}/\hat{\eta}^{\pi \pi}_{22} =
-\hat{\eta}^{\pi \pi}_{12}/\hat{\eta}^{\pi \pi}_{11}$ which follows from
first order perturbation theory in the strength of the two-pseudoscalar
interaction.  This approximation should introduce only a small error
since simply setting $\hat{\eta}^{\pi \pi}_{21}$ to 0 alters the final
$\lambda_1$ by less than a standard deviation.  As discussed earlier, if
$\lambda_1$ is determined from Eq.~(\ref{Sasym}) at large enough $t_g -
t_{\pi}$, the result is completely independent of $\hat{\eta}^{\pi
\pi}_{21}$.

\begin{figure}
\begin{center}
\vskip -5mm
\leavevmode
\epsfxsize=65mm
\epsfbox{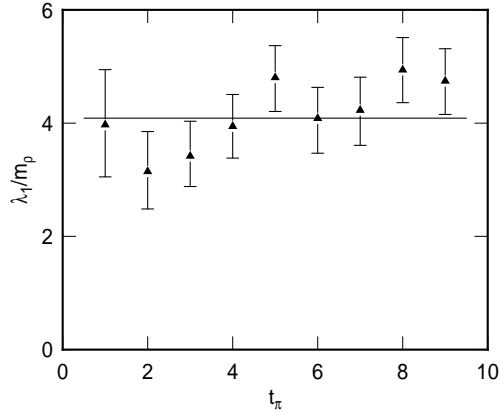}
\vskip -14mm
\end{center}
\caption{ 
$\lambda_1$ for $t_g - t_{\pi} = 0$ as a function of
$t_{\pi}$, compared with a fit on $t_g - t_{\pi} = 0$, $3 \le t_{\pi} \le 7$. 
}
\label{fig:g0t0}
\vskip -8mm
\end{figure}

\begin{figure}
\begin{center}
\vskip -5mm
\leavevmode
\epsfxsize=65mm
\epsfbox{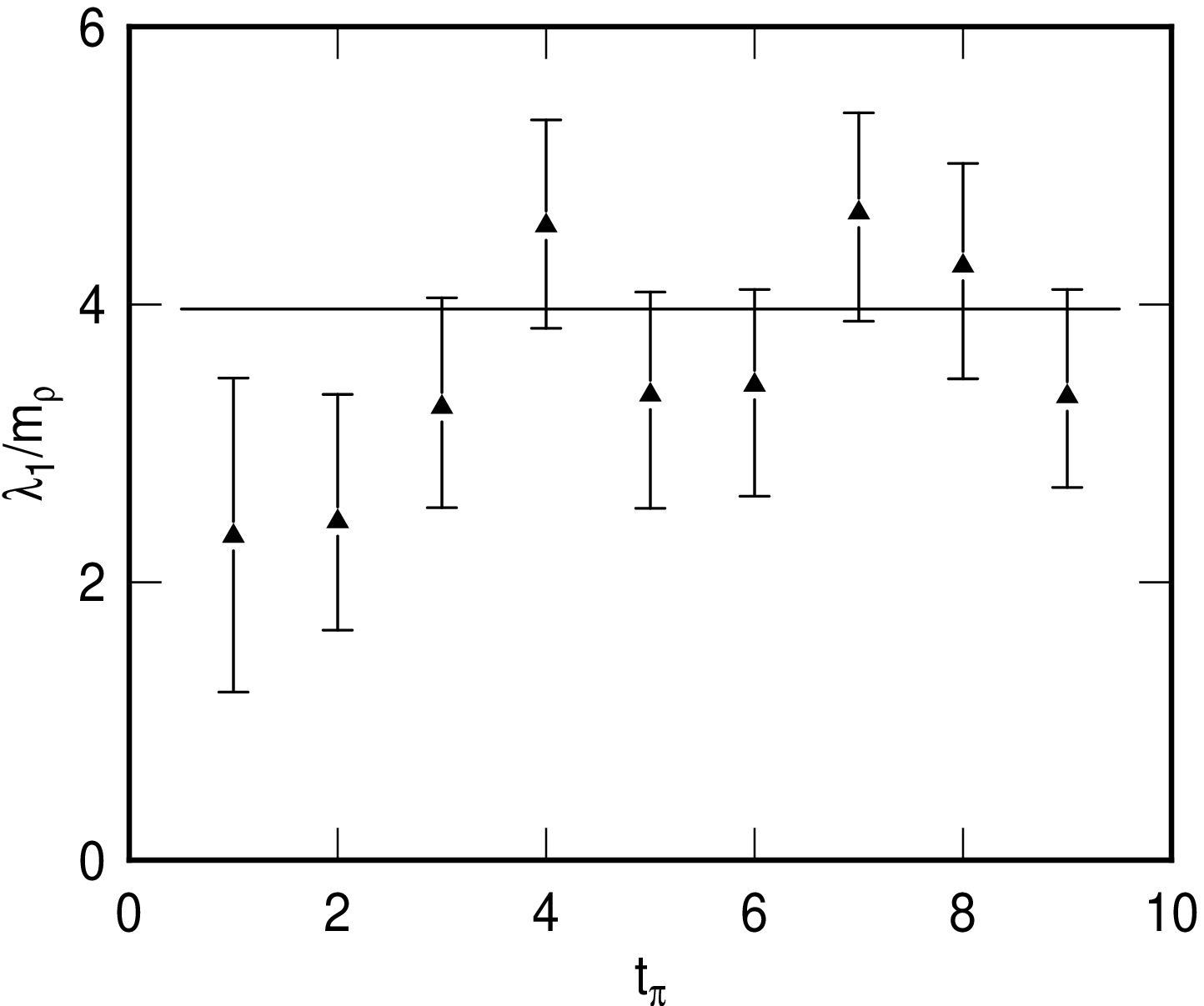}
\vskip -15mm
\end{center}
\caption{ 
$\lambda_1$ for $t_g - t_{\pi} = 1$ as a function of
$t_{\pi}$, compared with a fit on $t_g - t_{\pi} = 1$, $3 \le t_{\pi} \le 7$. 
}
\label{fig:g0t1}
\vskip -9mm
\end{figure}

\begin{figure}
\begin{center}
\vskip -5mm
\leavevmode
\epsfxsize=65mm
\epsfbox{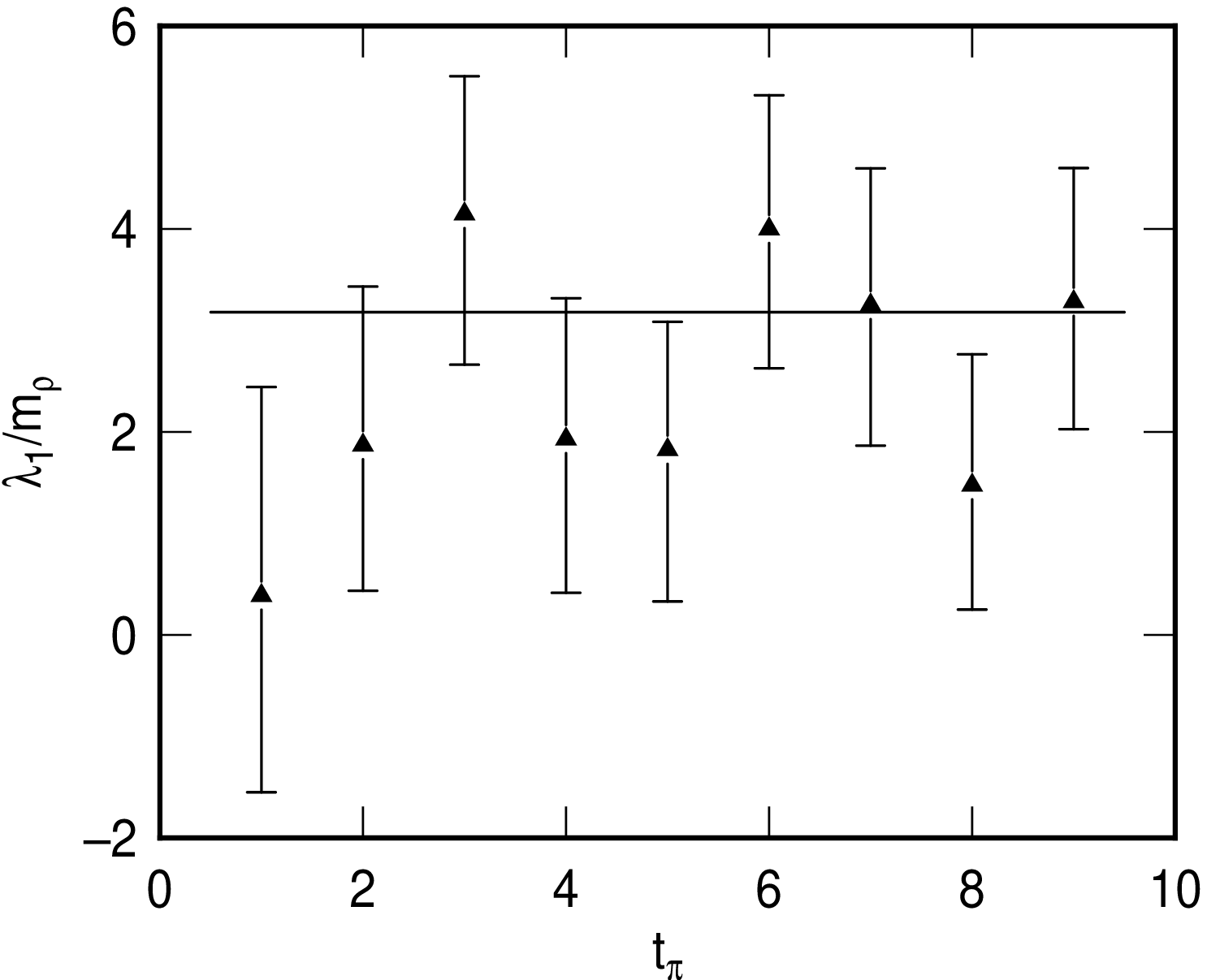}
\vskip -13mm
\end{center}
\caption{ 
$\lambda_1$ for $t_g - t_{\pi} = 2$ as a function of
$t_{\pi}$, compared with a fit on $t_g - t_{\pi} = 2$, $3 \le t_{\pi} \le 7$. 
}
\label{fig:g0t2}
\vskip -8mm
\end{figure}

\begin{figure}
\begin{center}
\vskip -5mm
\leavevmode
\epsfxsize=65mm
\epsfbox{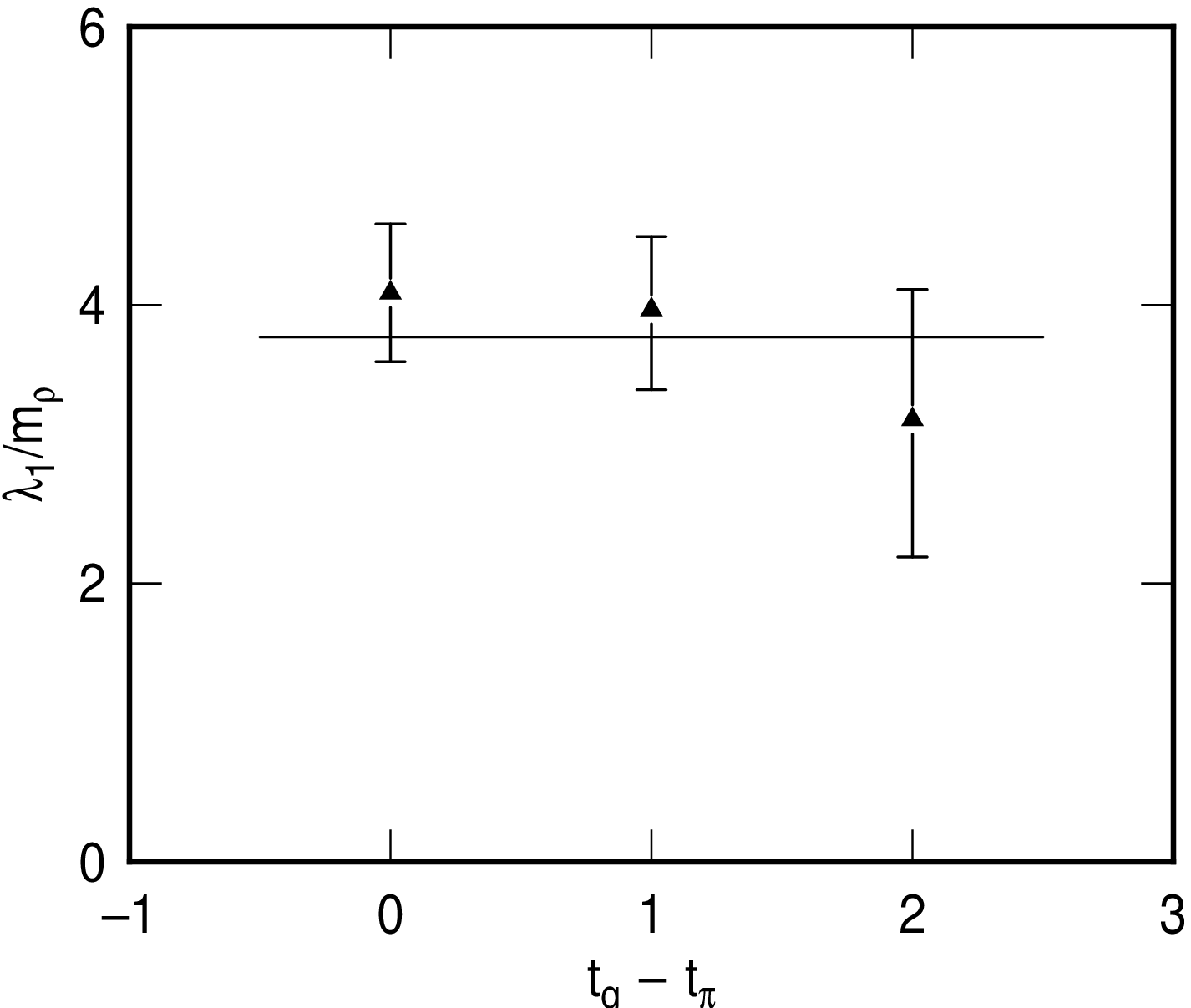}
\vskip -13mm
\end{center}
\caption{ 
$\lambda_1$ fitted on intervals $3 \le t_{\pi} \le 7$
with a single $t_g - t_{\pi}$ as a function of $t_g - t_{\pi}$,
compared with a fit on $1 \le t_g - t_{\pi} \le 2$, $3 \le t_{\pi} \le
7$.
}
\label{fig:g0t012}
\vskip -8mm
\end{figure}

\begin{figure}
\begin{center}
\vskip -5mm
\leavevmode
\epsfxsize=65mm
\epsfbox{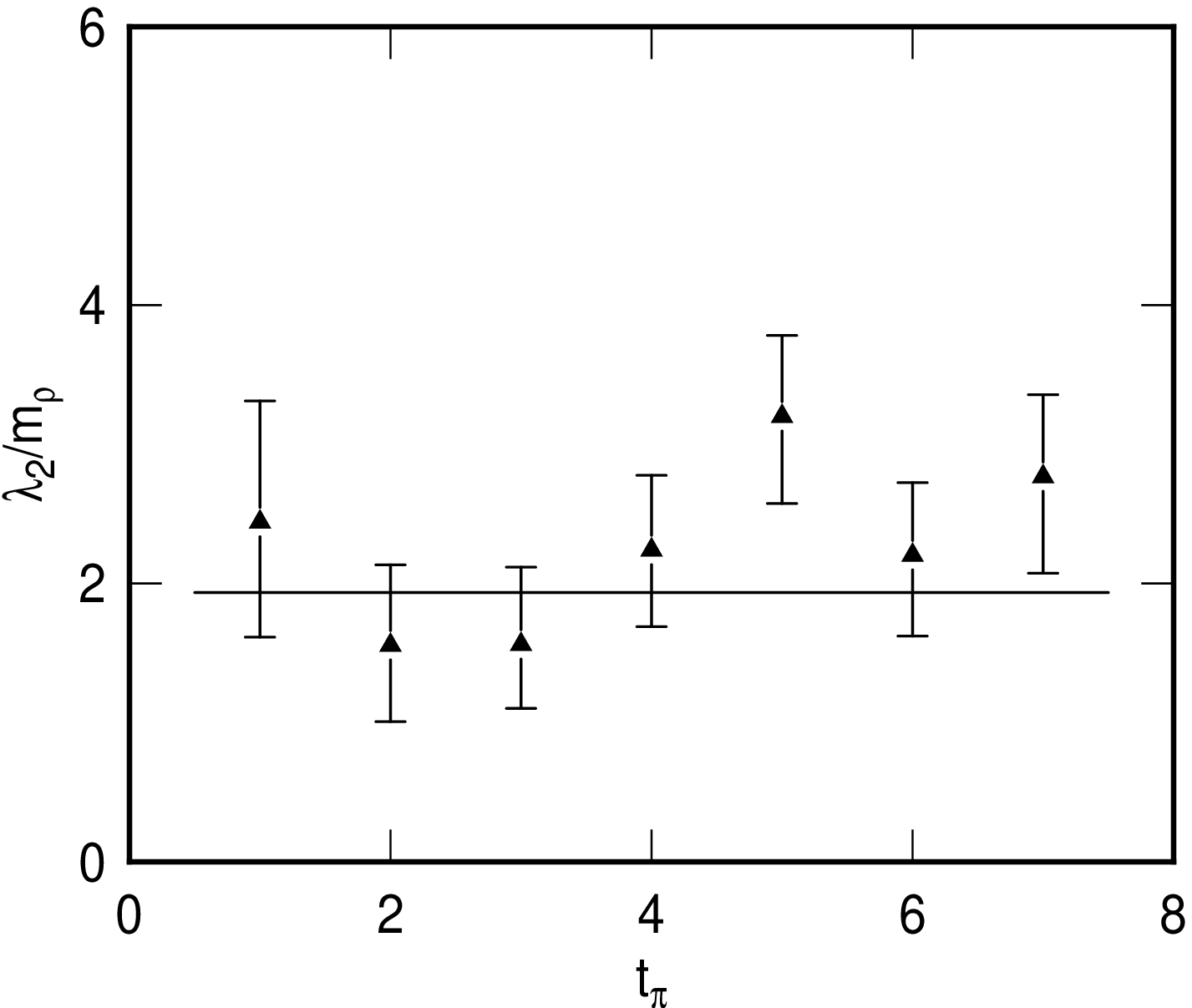}
\vskip -13mm
\end{center}
\caption{ 
$\lambda_2$ for $t_g - t_{\pi} = 0$ as a function of
$t_{\pi}$, compared with a fit on $t_g - t_{\pi} = 0$, $2 \le t_{\pi} \le 6$. 
}
\label{fig:g1t0}
\vskip -8mm
\end{figure}

\begin{figure}
\begin{center}
\vskip -5mm
\leavevmode
\epsfxsize=65mm
\epsfbox{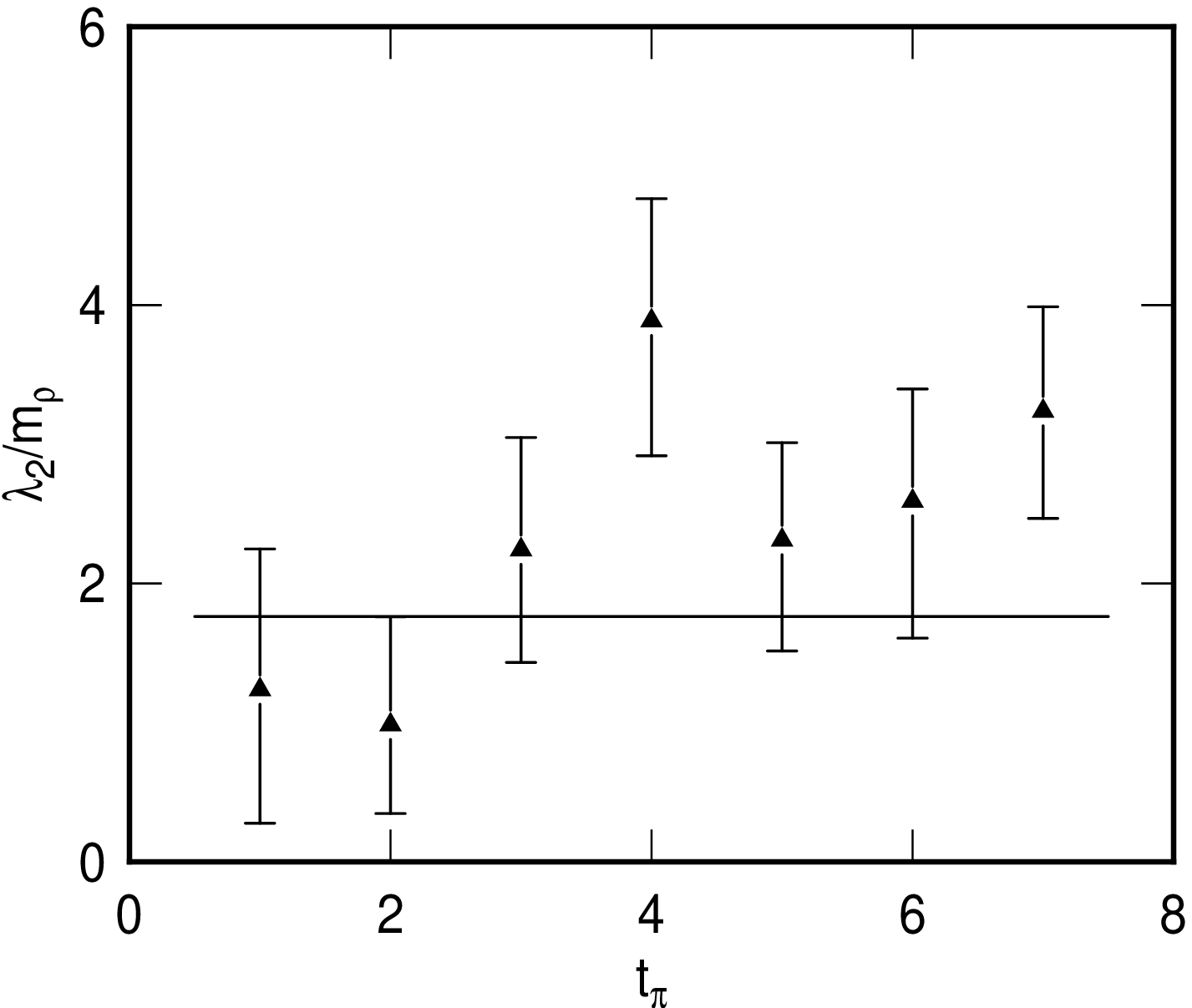}
\vskip -13mm
\end{center}
\caption{ 
$\lambda_2$ for $t_g - t_{\pi} = 1$ as a function of
$t_{\pi}$, compared with a fit on $t_g - t_{\pi} = 1$, $2 \le t_{\pi} \le 6$. 
}
\label{fig:g1t1}
\vskip -8mm
\end{figure}

\begin{figure}
\begin{center}
\vskip -5mm
\leavevmode
\epsfxsize=65mm
\epsfbox{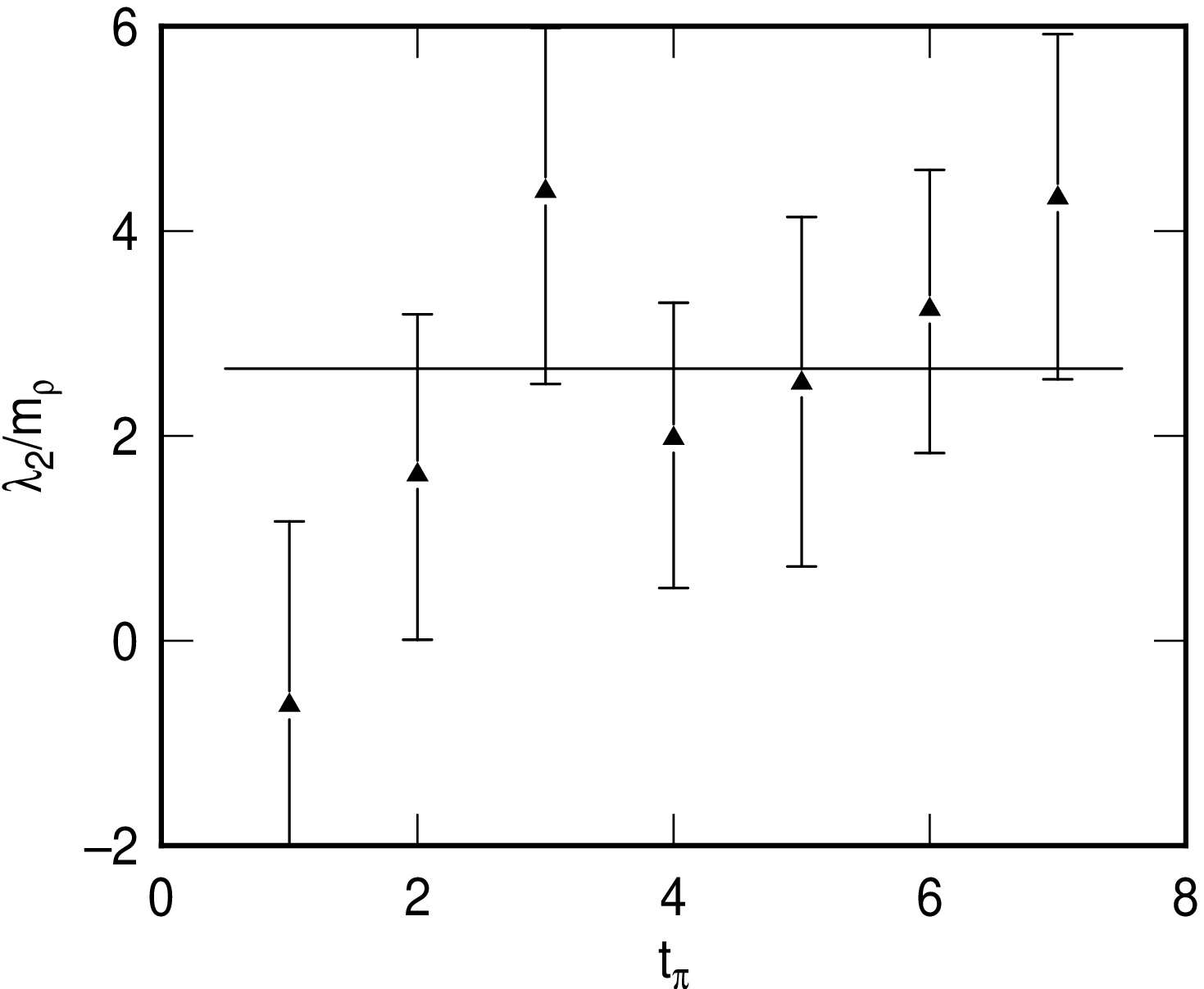}
\vskip -13mm
\end{center}
\caption{ 
$\lambda_2$ for $t_g - t_{\pi} = 2$ as a function of
$t_{\pi}$, compared with a fit on $t_g - t_{\pi} = 2$, $2 \le t_{\pi} \le 6$. 
}
\label{fig:g1t2}
\vskip -8mm
\end{figure}

\begin{figure}
\begin{center}
\vskip -5mm
\leavevmode
\epsfxsize=65mm
\epsfbox{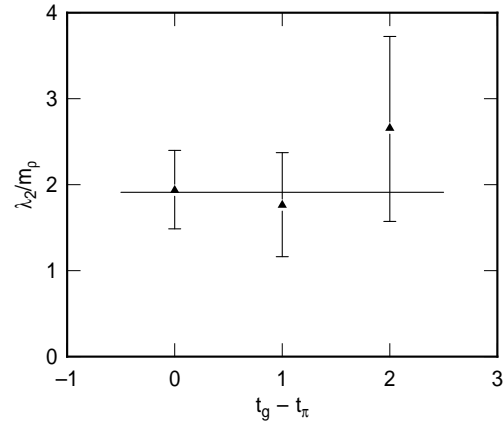}
\vskip -13mm
\end{center}
\caption{ 
$\lambda_1$ fitted on intervals $3 \le t_{\pi} \le 7$
with a single $t_g - t_{\pi}$ as a function of $t_g - t_{\pi}$,
compared with a fit on $0 \le t_g - t_{\pi} \le 1$, $2 \le t_{\pi} \le
6$.
}
\label{fig:g1t012}
\vskip -8mm
\end{figure}

From our 10500 configuration ensemble on a $16^3 \times 24$ lattice, we
evaluated $S_1$ and $S_2$ for glueball decay on mass shell at $\kappa$
of 0.1650 and 0.1675, respectively. We obtained statistically
significant results for $0 \le t_g - t_{\pi} \le 2$ with $0 \le t_{\pi}
\le 9$ for $S_1$ and with $0 \le t_{\pi} \le 7$ for $S_2$. 
At each point within this range we then determined
effective $\lambda_i$ using Eq.~(\ref{Sasym}).  We found $\lambda_1$ and
$\lambda_2$ statistically consistent with being constant for $t_{\pi}
\ge 3$ and $t_{\pi} \ge 2$, respectively, and all values of $t_g -
t_{\pi}$.  
Figures~\ref{fig:g0t0} 
shows effective
$\lambda_1$ in units of the $\rho$ mass as a function of $t_{\pi}$ for
$t_g - t_{\pi} = 0$ in comparison a fit with $3 \le t_{\pi} \le 7$,
$t_g - t_{\pi} = 0$. Figures~\ref{fig:g0t1} and \ref{fig:g0t2}
show corresponding data for $t_g - t_{\pi}$ of 1 and 2, respectively.   
Figure~\ref{fig:g0t012} shows fitted values of
$\lambda_1$ on the interval $3 \le t_{\pi} \le 7$ for fixed $t_g -
t_{\pi}$ of 0, 1 or 2.  
Figures~\ref{fig:g1t0} 
shows effective
$\lambda_2$ in units of the $\rho$ mass as a function of $t_{\pi}$ for
$t_g - t_{\pi} = 0$ in comparison a fit with $2 \le t_{\pi} \le 6$,
$t_g - t_{\pi} = 0$. Figures~\ref{fig:g1t1} and \ref{fig:g1t2}
show corresponding data for $t_g - t_{\pi}$ of 1 and 2, respectively.   
Figure~\ref{fig:g1t012} shows fitted values of
$\lambda_1$ on the interval $2 \le t_{\pi} \le 6$ for fixed $t_g -
t_{\pi}$ of 0, 1 or 2.  

To extract final values of $\lambda_i$, we tried fits to all rectangular
intervals of data including at least 4 values of $t_{\pi}$ and at least
2 values of $t_g - t_{\pi}$. For each $\lambda_i$ we chose the fit
giving the lowest value of $\chi^2$ per degree of freedom. The window
determined in this way for $\lambda_1$ is $3 \le t_{\pi} \le 7$ with $1
\le t_g - t_{\pi} \le 2$, and for $\lambda_2$ is $2 \le t_{\pi} \le 6$
with $0 \le t_g - t_{\pi} \le 1$.  The horizontal lines in
Figures~\ref{fig:g0t012} and \ref{fig:g1t012} shows the final value of
$\lambda_1$ and $\lambda_2$ respectively.  Over the full collection of
windows we examined, the fitted results varied from our final results by
at most 1 standard deviation. We believe our best fits provide
reasonable estimates of the asymptotic coefficients in
Eq.~(\ref{Sasym}).

So far our discussion has been restricted to QCD with u, d and s quark
masses degenerate. An expansion to first order in the quark mass matrix
taken around some relatively heavy SU(3) symmetric point gives glueball
decay couplings for $\pi$'s, K's and the $\eta$ which are a common
linear function of each meson's average quark mass. Since meson masses
squared are also nearly a linear function of average quark mass, the
decay couplings are a linear function of meson masses squared. Thus from
a linear fit to our predictions for decay couplings as a function of
pseudoscalar mass squared at unphysical degenerate values of quark
masses we can extrapolate decay couplings for physical nondegenerate
values of quark masses.  From this linear fit a prediction can also be
made for the decay coupling of the scalar glueball to $\eta + \eta'$, if
we ignore the contribution to the decay from the process in which the
$\eta$ quark and antiquark are connected to each other by one propagator
and the $\eta'$ quark and antiquark are connected to each other by a
second propagator.

Figure~\ref{fig:lambdas} shows predicted coupling constants as a function
of predicted meson mass squared along with linear extrapolations of the
predicted values to the physical $\pi$, K and $\eta$ masses, in
comparison to observed decay couplings\cite{Lind} for decays of
$f_J(1710)$ to pairs of $\pi$'s, K's and $\eta$'s.  Masses and decay
constants are shown in units of the $\rho$ mass. Our predicted width for
the scalar glueball decay to $\eta + \eta'$ is 6(3) MeV. For the ratio
$\lambda_{\eta \eta'} / \lambda_{\eta \eta}$ we get 0.52(13). We predict
a total width for glueball decay to pseudoscalar pairs of 108(29) MeV,
in comparison to 99(15) MeV for $f_J(1710)$.

One of us (D.W.) is grateful to S.\ Lindenbaum, R.\ Longacre, S.\ Sharpe
and the participants in Gluonium 95 for valuable conversations.

\end{document}